\documentclass[%
 reprint,
 amsmath,amssymb,
 aip,
]{revtex4-1}

\usepackage{graphicx}
\usepackage{dcolumn}
\usepackage{bm}

\makeatletter 
\newcommand*{\rom}[1]{\expandafter\@slowromancap\romannumeral #1@} 
\makeatother

\begin{document}

\preprint{APS/123-QED}

\title{Complex band structure and electronic transmission}

\author{Anders Jensen}
\affiliation{Nano-Science Center and Department of Chemistry, University of Copenhagen, 2100 Copenhagen \O, Denmark}
\author{Mikkel Strange}
\affiliation{Nano-Science Center and Department of Chemistry, University of Copenhagen, 2100 Copenhagen \O, Denmark}
\affiliation{Department of Physics, Technical University of Denmark, 2800 Kongens Lyngby, Denmark}
\author{S\o ren Smidstrup}
\author{Kurt Stokbro}
\affiliation{QuantumWise A/S, Fruebjergvej 3, Box 4, 2100 Copenhagen, Denmark}
\author{Gemma C. Solomon}
\email{gsolomon@nano.ku.dk}
\affiliation{Nano-Science Center and Department of Chemistry, University of Copenhagen, 2100
Copenhagen \O, Denmark}
\author{Matthew G. Reuter}
\email{matthew.reuter@stonybrook.edu}
\affiliation{Department of Applied Mathematics \& Statistics and Institute for Advanced Computational Science, Stony Brook University, Stony Brook, New York 11794, United States}

\date{\today}

\begin{abstract}  
The function of nano-scale devices critically depends on the choice of materials. For electron transport junctions it is natural to characterize the materials by their conductance length dependence, $\beta$. Theoretical estimations of $\beta$ are made employing two primary theories: complex band structure and DFT-NEGF Landauer transport. Both reveal information on $\beta$ of individual states; i.e. complex Bloch waves and transmission eigenchannels, respectively. However, it is unclear how the $\beta$-values of the two approaches compare. Here, we present calculations of decay constants for the two most conductive states as determined by complex band structure and standard DFT-NEGF transport calculations for two molecular and one semi-conductor junctions. Despite the different nature of the two methods, we find strong agreement of the calculated decay constants for the molecular junctions while the semi-conductor junction shows some discrepancies. The results presented here provide a template for studying the intrinsic, channel resolved length dependence of the junction through complex band structure of the central material in the heterogeneous nano-scale junction.    
\end{abstract}

\maketitle


\section{Introduction}
\label{sec:introduction}
In the design of nanoscale electronic components, it is crucial to understand and control the behavior of materials. This is true whether the objective is to avoid leakage currents in transistors or to find the optimal material for a thermoelectric device. Today, many modern electronic components are on the scale of nanometers and thus in the coherent electron tunneling regime, where the conductance of a material decays exponentially with its length. The conductance of a repeat material, whether it be a molecule or a solid state system, has a characteristic length dependence $\beta$ and can expressed as \cite{Magoga1996}
\begin{align}
G = G_c e^{- \beta L},
\end{align}
where $G_c$ and $L$ are the contact conductance and length of the material, respectively. Experimental and theoretical efforts characterize materials based on their $\beta$-value. The theoretical studies often estimate $\beta$ from the electron transmission which is proportional to conductance at low bias. Two main approaches have been applied to this end, one developed in chemistry and the other in solid state physics.

1) In chemistry, experimental and theoretical studies of $\beta$ have measured and modeled electron conductance in both electron transfer \cite{Gray1996} and electron transport through single molecules \cite{Salomon2003,tomfohr-1542-2004}. A single molecule junction consists of two electrodes connected by a molecular bridge. While it is clear that $\beta$ is dominated by the backbone of the molecular bridge in these systems, band alignment (as influenced by the nature of the electrodes and molecular binding groups) and other effects can modify $\beta$ \cite{peng2009conductance}. 
The relevant electronic states are the scattering states or equivalently the transmission eigenchannels \citep{paulsson2007transmission} extending across the whole of junction including the electrodes. 

2) In solid-state physics, where band structure is the language of choice to describe the electronic structure of periodic solids, we find another method to study $\beta$: complex band structure \cite{heine-1689-1965,Kohn1959,reuter2016unified}. Complex band structure has the familiar real bands of conventional band structure, but also includes complex bands that describe Bloch states with complex wave vectors. The complex bands describe the evanescent states of periodic materials, and have been utilized in a variety of applications \cite{reuter2016unified}, including Fermi level alignment \cite{Tersoff1984,tomfohr-245105-2002}, metal-induced gap states \cite{tomfohr-245105-2002}, edge/impurity states \cite{Park2013} and conductance length dependence  \cite{tomfohr-245105-2002,tomfohr-59-2002,picaud-3731-2003,fagas-268-2004}. In contrast with transport theory, only a homogeneous solid or molecule is studied in complex band structure, corresponding to the backbone of the bridge material in a junction. The relevant electronic states here are the Bloch states with complex wave vectors.

Naturally this leaves the question, how are the complex Bloch states related to the transmission eigenchannels? In the studies mentioned above, estimates of the total transmission/conductance length dependence from complex band structure are based on the slowest decaying Bloch state at a given energy. This assumes only one contributing channel and that the length dependence of this channel transmission is equal to the slowest decaying Bloch state. For the longer molecules, these assumptions appear valid in the systems studied. However, for shorter molecules or more complex systems these assumptions might not be true. Such complexities could arise from multiple contributing channels or destructive interference. It remains an open question as to how these two methods compare for systems with these attributes.

In this paper we explore the relationship between the decay of the individual complex Bloch states and the transmission length dependence of scattering states/transmission eigenchannels in detail. The paper proceeds as follows. We first build a tight-binding model junction that illustrates how the scattering states can be built from Bloch states. Second, we turn to a more realistic atomistic description of three junctions designed to answer the questions of multiple contributing channels and destructive interference. To achieve this, we compare the two slowest decaying Bloch states with the transmission of the two transmission eigenchannels that contribute the most to the transmission (proportional to the conductance at small bias). We conclude with a summary of our findings.  


\section{Model system} 
\label{cbs}
In this section we illustrate the connection between the two methods: complex band structure and scattering theory. We first build a tight-binding model to show the nature of the complex band structure and its corresponding Bloch states. The model is then expanded into a model junction with a simple square potential barrier. This illustrates how the scattering states are built from different Bloch states and how the decay inside the barrier can be used to calculate the transmission and the transmission length dependence. Although this model and the connection between complex Bloch states and transmission length dependence has previously been shown by Tomfohr and Sankey \cite{tomfohr-245105-2002}, we extend the presentation by illustrating how the model can be directly expanded into a model junction and give visual representations that illustrate the connection between the two methods and their relevant states. 
\subsection{Tight-binding model}
\label{tbsec}
\begin{figure}[b]
 	\centering
 	\includegraphics[width=1\linewidth]{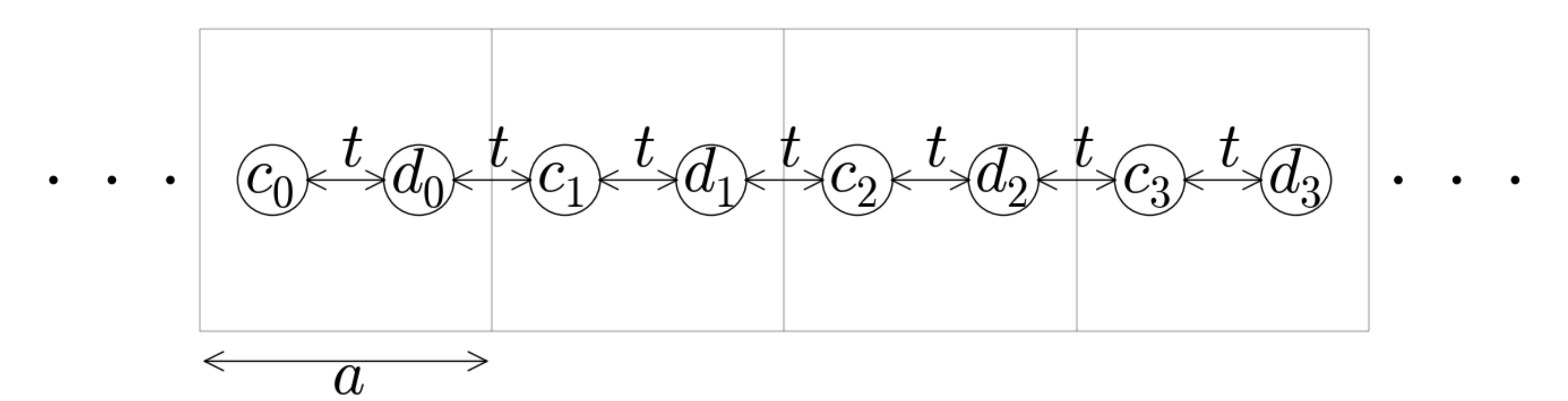}
 	\caption{An illustration of the one dimensional tight-binding model employed for an infinite alternating chain of two different sites. The parameters $a$, $t$, $c_i$ and $d_i$ represent the unit cell length, hopping element and site labels for the two sites in unit cell $i$, respectively.}
 	\label{tbmodel}
\end{figure}

We begin with the derivation of the complex band structure for a one-dimensional tight-binding model of an infinite alternating chain of two different sites, see Figure \ref{tbmodel}.
The Hamiltonian for the model is
\begin{align}
H &= \sum_i \left[ \epsilon_c c_i^{\dagger} c_i + \epsilon_d d_i^{\dagger} d_i + t \left( c_i^{\dagger} d_i + c_{i+1}^{\dagger} d_{i} + H.c. \right) \right],
\end{align}
where $c_i$ and $d_i$ are the two sites in the unit cell of index $i$. $\epsilon_c$ and $\epsilon_d$ are the onsite energies on sites $c$ and $d$, respectively, and $t$ is the hopping element. As the model is infinite and periodic, it is convenient to switch the description from real space to Fourier space, which allows us to map the Hamiltonian into a 2 $\times$ 2 matrix
\begin{align}
H_k = \left(\begin{matrix}
\epsilon_c & 2 t \cos\left(\frac{k a}{2}\right) \\
2 t \cos\left(\frac{k a}{2}\right) & \epsilon_d
\end{matrix} \right),
\end{align}
where $k$ is the wave number and $a$ is the length of the unit cell. The Fourier Hamiltonian has two eigenenergies (forming bands when considered over all $k$),
\begin{subequations}
\begin{align} \label{eigen}
\epsilon_{k}^{\pm} = \frac{\epsilon_c + \epsilon_d}{2} \pm \frac{1}{2} \sqrt{16 t^2 \cos \left(\frac{k a}{2} \right)^2+(\epsilon_c - \epsilon_d)^2},
\end{align}
and the eigenstates are of the Bloch form
\begin{align}
\psi_{k}^{(+) \dagger} = v_c^{+}(k) c_k^{\dagger} + v_d^{+}(k) d_k^{\dagger},\\
\psi_{k}^{(-) \dagger} = v_c^{-}(k) c_k^{\dagger} + v_d^{-}(k) d_k^{\dagger},
\end{align}
\end{subequations}
where $v_c(k)$ and $v_c(k)$ are discrete periodic functions with a period of $a$. $ c_k^{\dagger}$ and $d_k^{\dagger}$ are creators of plane waves. 

\begin{figure}[]
 	\centering
 	\includegraphics[width=0.99\linewidth]{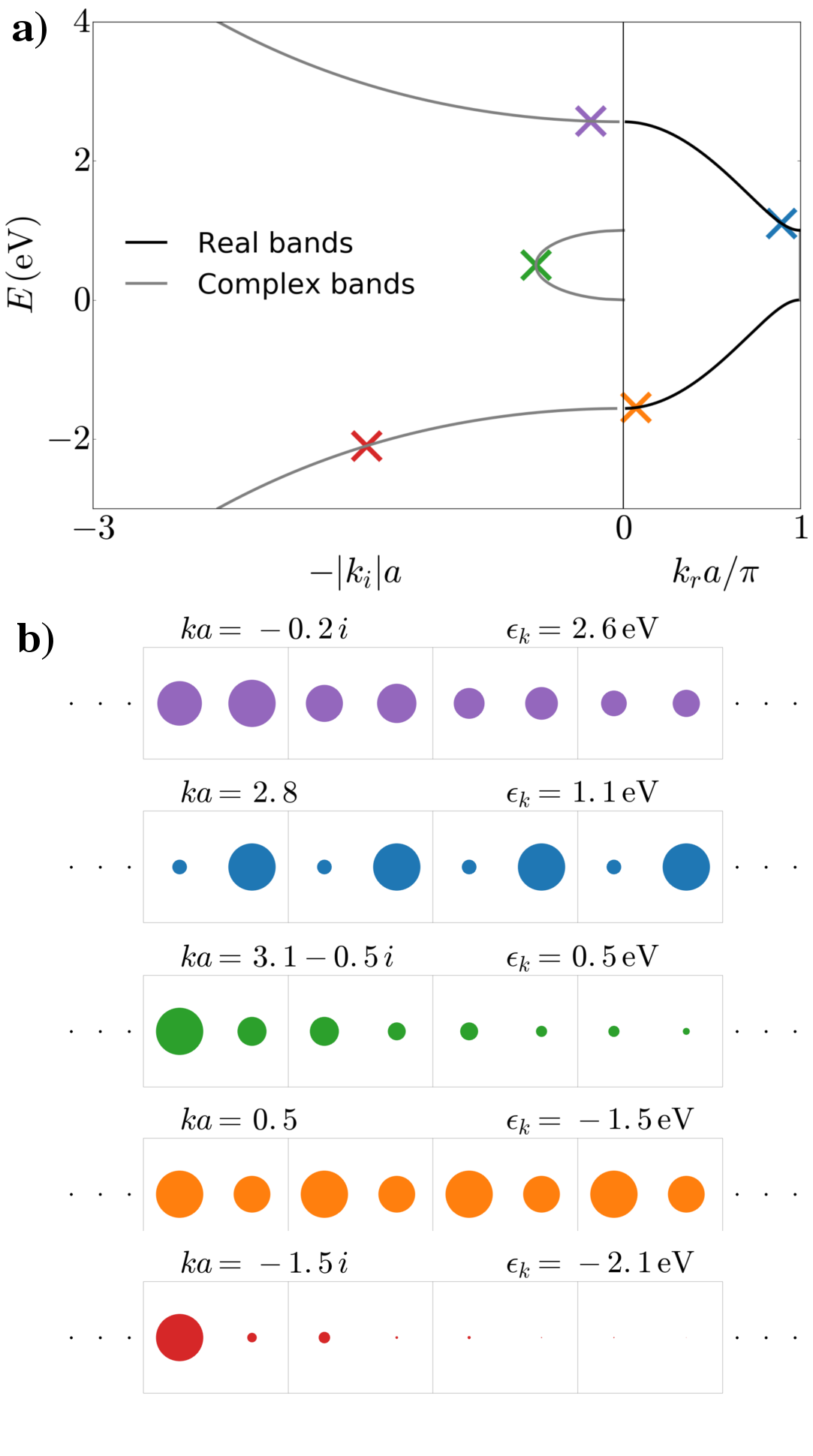}\\ 		
 	\caption{\textbf{a)} Complex band structure for a two site infinite model. The two black lines are the real bands and the three grey lines are complex bands, of which two are purely imaginary. \textbf{b)} The amplitude of the eigenstates at five energies marked with a colored cross. }
 	\label{cbs_model}
\end{figure}

In conventional band structure calculations, periodic boundary conditions are enforced. However, to calculate the complex bands, this constraint is lifted, giving solutions with complex $k$ that are local in space and sometimes referred to as edge states. The strategy for finding these states is to isolate $k$ in Equation \ref{eigen},

\begin{align}
k = \frac{2}{a} \cos^{-1}\left(\frac{1}{2 t} \sqrt{(E - \epsilon_c)(E - \epsilon_d)}\right) + \frac{4 \pi n}{a},
\end{align}
where $n \in \mathbb{Z}$. All solutions to this equation with real eigenenergies $\epsilon_k$ are also solutions to the time-independent Schr\"odinger equation. As an example, we choose the model parameters $t = -1 \, \text{eV}$, $\epsilon_c = 0 \, \text{eV}$, $\epsilon_d = 1 \, \text{eV}$, $a = 1 \, \text{\AA}$ and plot the solutions and examples of eigenstates (squared) in Figure \ref{cbs_model}, where we have defined the real and imaginary part of the wave number, 
\begin{align}
k = k_r + i k_i.
\end{align}
The complex band structure consists of two real bands, with complex bands outside of the real bands, see Figure \ref{cbs_model} a). The complex band in the middle connects the two real bands whereas the two outer, purely imaginary bands continue to grow in magnitude with larger $k_i$ values. The band gap is 1 eV, equivalent to the difference between the on-site energies of sites $c$ and $d$.

In Figure \ref{cbs_model} b), the top (purple) eigenstate is high in energy (2.6 eV) and has a small $k_i$, which gives a slow decay in the amplitude. The amplitude is larger on the $d$ site than the $c$ site, which is a characteristic shared with states in the conduction band (blue). Remember $d$ sites have a higher on-site energy. The second (blue) eigenstate has $k_i=0$ and thus no decay in amplitude. The third (green) eigenstate decays faster than the first eigenstate because of the magnitude of $k_i$ is larger. The real part giving the phase is obscured because we are plotting the amplitude. The fourth (orange) eigenstate is in the valence band and is mostly located on the $c$ site, as it has the lower on-site energy. Again because $k$ is purely real there is no decay. Finally the fifth (red) eigenstate has the largest $k_i$ in magnitude and consequently the fastest decay. Generally, the amplitude on site $c$ and $d$ for any given state is controlled by which real band that state is located on or, if it is located on a complex band, which real band it is closest to. We emphasize that the rate of decay (or lack thereof) is controlled by the magnitude of $k_i$.

\subsection{Expansion to model junction}
\begin{figure}[]
 	\centering
	\includegraphics[width=0.99\linewidth]{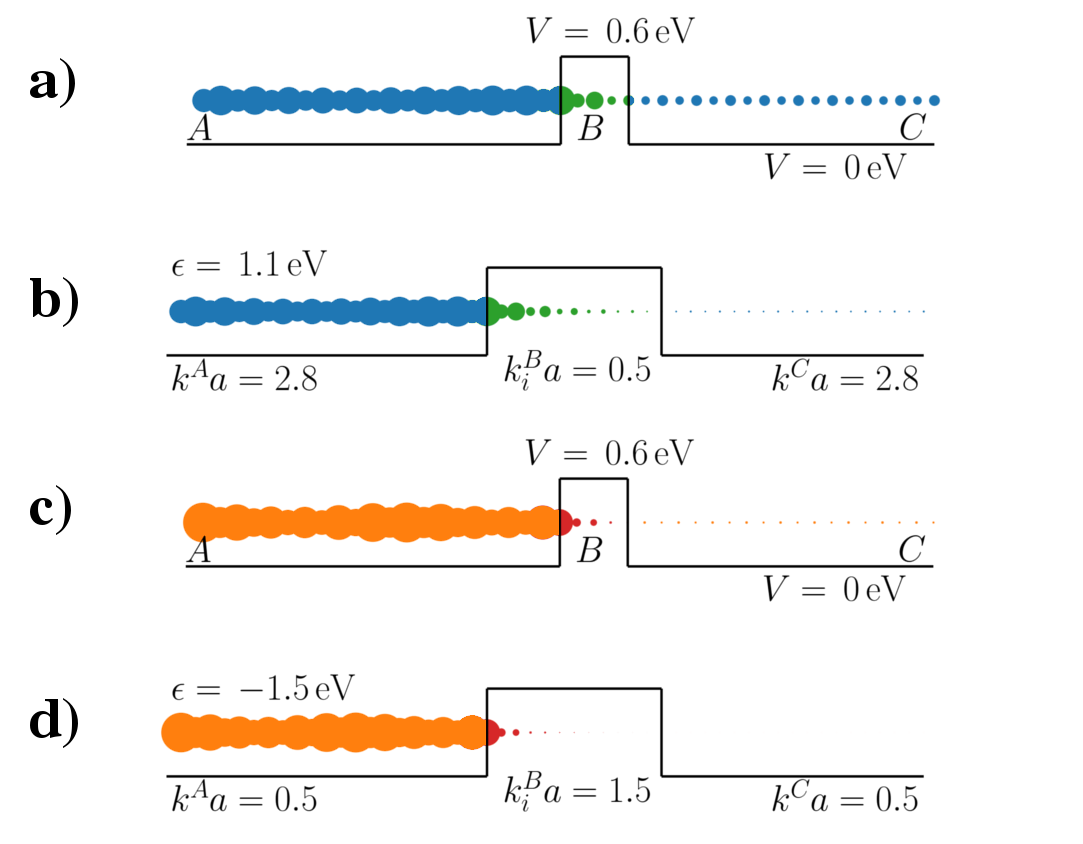}		
 	\caption{Four model junctions and their eigenstates. In region \textit{B} the potential is shifted $0.6$ eV, meaning that the corresponding complex band structure would be the same as in Figure \ref{cbs_model} but with all bands shifted $0.6$ eV up. \textbf{a)} and \textbf{b)} only differ in the length of the barrier: 2 and 6 unit cells, respectively. Likewise for \textbf{c)} and \textbf{d)}. }   
 	\label{junction_model}
\end{figure}

The model of the previous section is now expanded to represent a junction by adding a simple rectangular potential barrier. We use an approximate method which allows us to clearly connect the Bloch states of the infinite model with scattering states of the model junction and transmission through it. We divide the infinite chain into three regions $A$, $B$ and $C$. We consider $A$ and $C$  as two semi-infinite electrodes, whereas the region $B$ is the finite central region where the on-site energies have been shifted by a constant potential (the rectangular barrier). The total wave function of the junction now has the form
\begin{align} \label{cwf}
\Psi &= p_{k^{A}} \psi_{k^{A}} + p_{-k^{A}} \psi_{-k^{A}} \\ &+ p_{k^{B}} \psi_{k^{B}}  + p_{-k^{B}} \psi_{-k^{B}} \\ &+ p_{k^{C}} \psi_{k^{C}} + p_{-k^{C}} \psi_{-k^{C}}.
\end{align}
Here $p_{\pm k^n}$ is a $\pm k^n$-dependent amplitude in region $n$. In regions $A$ and $C$ we enforce periodic boundary conditions, because they are semi-infinite regions corresponding to leads in a junction. Consequently only real $k$ solutions are allowed. The central region $B$ is finite and can have different boundary conditions in the directions of regions $A$ and $C$, thus complex $k$ solutions are also allowed. Four examples are shown in Figure \ref{junction_model}. 
The model parameters are the same as in Section \ref{tbsec} and the potential in region $B$ is shifted to $V = 0.6$ eV. The energy of the electron crossing the junction is $\epsilon = 1.1$ eV for a) and b) and  $\epsilon = -1.5$ eV for c) and d). That is, an electron from the conduction and valence band of region $A$, respectively. For a) and c) the length of the barrier is 2 unit cells while for b) and d) the length is 6 unit cells.  The boundary conditions require that the amplitudes, $p_{\pm k^n}$, of the wave functions match at the interfaces between the three regions. To find the amplitudes using the boundary conditions we set $p_{k^{A}}=1$ and $p_{-k^{C}}=0$. Here, because our model is discrete, this matching is made in the unit cell before the barrier. The colors correspond to the colors of the eigenstates in Figure \ref{cbs_model}. 

Naturally we observe that the wave function in region $A$ consist of both a right going ($+k$) and and left going ($-k$) part. In region $C$ we see only a right going wave since we set $p_{-k^{C}}=0$. In region $B$ of a) and c), we see a difference in the rate of decay even though the barrier height is the same. The reason is that $k^B_i$ depends on both the barrier height and the energy of the tunneling electron.  

From the boundary conditions we can derive an expression for the transmission of the composite wave function in Equation \ref{cwf},
\begin{align} \label{modeleq}
T = \frac{|p_{k^C}|^2}{|p_{k^A}|^2} \propto e^{-2 |k^B_i| L}.
\end{align}
Thus we find the corresponding decay constant for the model system is simply $\beta = 2 |k^B_i|$. This result was also found by Tomfohr and Sankey, albeit with a different approach \cite{tomfohr-245105-2002}. Note that our model is an approximation and does not entirely reproduce the standard Green's function method transmission, see SI.

In Section \ref{tbsec}, we illustrated the nature of the complex band structure and the corresponding Bloch states. Now we have also shown a model junction where the scattering states can be described as a linear combination of the Bloch states from the three regions in the junction. Finally we saw that the transmission length dependence $\beta$ could be determined from the decay of the Bloch states in the central region.    

In our model we only considered one scattering state, but in a real system there are many and they all have different features. This leaves us with the questions: how well does the complex band structure of real systems describe the transmission length dependence of the individual transmission eigenchannels? And, how do features such as destructive interference appear in the complex band structure? To find out, we turn to an atomistic description of realistic systems.  

\section{First principles calculations} 
\label{fp}

\begin{figure}[h!]
 	\centering
	\includegraphics[width=0.99\linewidth]{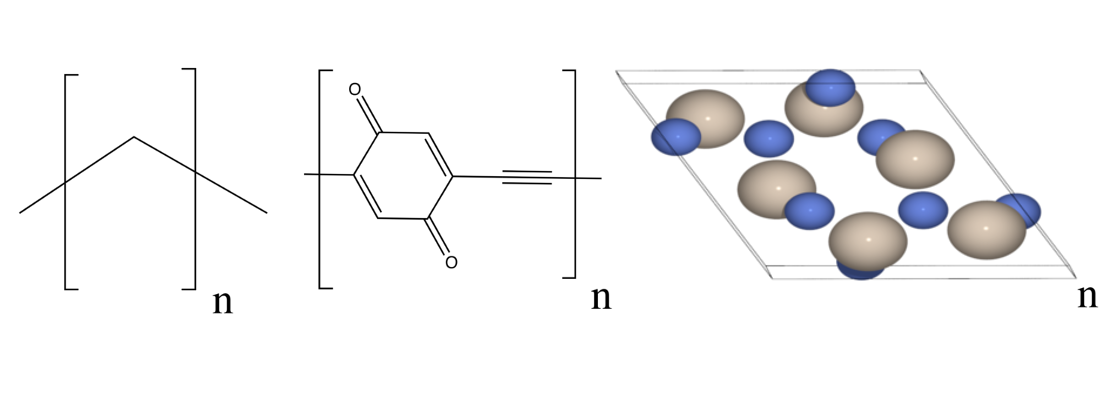}
 	\caption{Chemical structure of the alkane chain (left) and quinone chain (middle). Unit cell of Si$_3$N$_4$ (right).}
 	\label{chem_structure}
\end{figure}

Here we present a study on the relationship between the decay of Bloch states and the transmission eigenvalue length dependence. We performed transport calculations on three types of junctions: 1) Alkane chains with gold electrodes 2) Quinone chain with gold electrodes and 3) Si-Si$_3$N$_4$-Si semiconductor junctions. Furthermore, we made complex band structure calculations on the bridge materials, see Figure \ref{chem_structure} for chemical structures of the molecular chains and the unit cell of Si$_3$N$_4$. 

The junctions were chosen based on the criteria that they varied in the size of the band gap and the absence or presence of either destructive interference or multiple contributing transmission eigenchannels. 

For each of the three types of junctions we compare the length dependence of the two  dominant transmission eigenchannels $\beta_n$, $n = 1,2$, to the complex bands of the bridge material. The specifics of our approach are described below.

\subsection{Method}

To calculate the length dependence for two individual eigenchannels we create five junctions of different lengths for each of the three systems. 
\begin{itemize}
\item Alkane chain, n = 4,6,8,10,12
\item Quinone chain, n = 2,3,4,5,6
\item Si$_4$N$_3$ n, = 4,6,8,10,12 
\end{itemize}
These lengths were chosen to remove interface effects while maintaining coherent transport. The molecular chains were first terminated with saturated thiols and optimized in gas phase, then the hydrogen atoms at both ends were removed to create stable S-Au bonds in the gold junction. The molecules were placed at 2 \AA \ for the alkane chains and at 2.1 \AA \ for the quinone chain with respect to the gold surface at an FCC hollow site (the minimum energy distance as determined by a series of single point energy calculations at varying Au-S separation). The Si-Si$_3$N$_4$-Si junction is based on the $\beta$-Si3N4(0001)/Si(111) interface; see the work of Yang \textit{et al.} \cite{Yang2009}. The central region, consisting of the Si$_{3}$N$_{4}$ and a few layers of silicon on either side, was optimized while the electrodes were kept fixed. 

Next the transmission eigenvalues were calculated for the two channels with the largest contribution to the total transmission, $T_1$ and $T_2$ \cite{paulsson2007transmission}, for each of the five system lengths; see Figure \ref{method} a) for the quinone junction. 
 
All calculations were performed using density functional theory (DFT) with Atomistix ToolKit (ATK) version 2016.3 using the PBE exchange-correlation functional and a double zeta polarized basis set for all atoms \cite{brandbyge2002,QW}. The transport calculations were performed using the non-equilibrium Green's function Landauer approach. The transmission eigenchannels were averaged over 5 $\times$ 5 Monckhorst-Pack $k$-point mesh in the transverse direction. Note that the order of the channels must be unchanged across lengths and $k$-points for this analysis to work. However, changes could be accounted for by comparing eigenchannels across energies.

The complex band structure was also calculated using DFT in the ATK software. The method implemented to compute the complex band structure follows the companion matrix method of Chang and Schulman \cite{Chang1982}. The unit cells for the complex band structure calculations were optimized using variable unit cell dimensions. 

\subsection{Results \& Discussion}

\begin{figure}[h!]
 	\centering
	\includegraphics[width=0.99\linewidth]{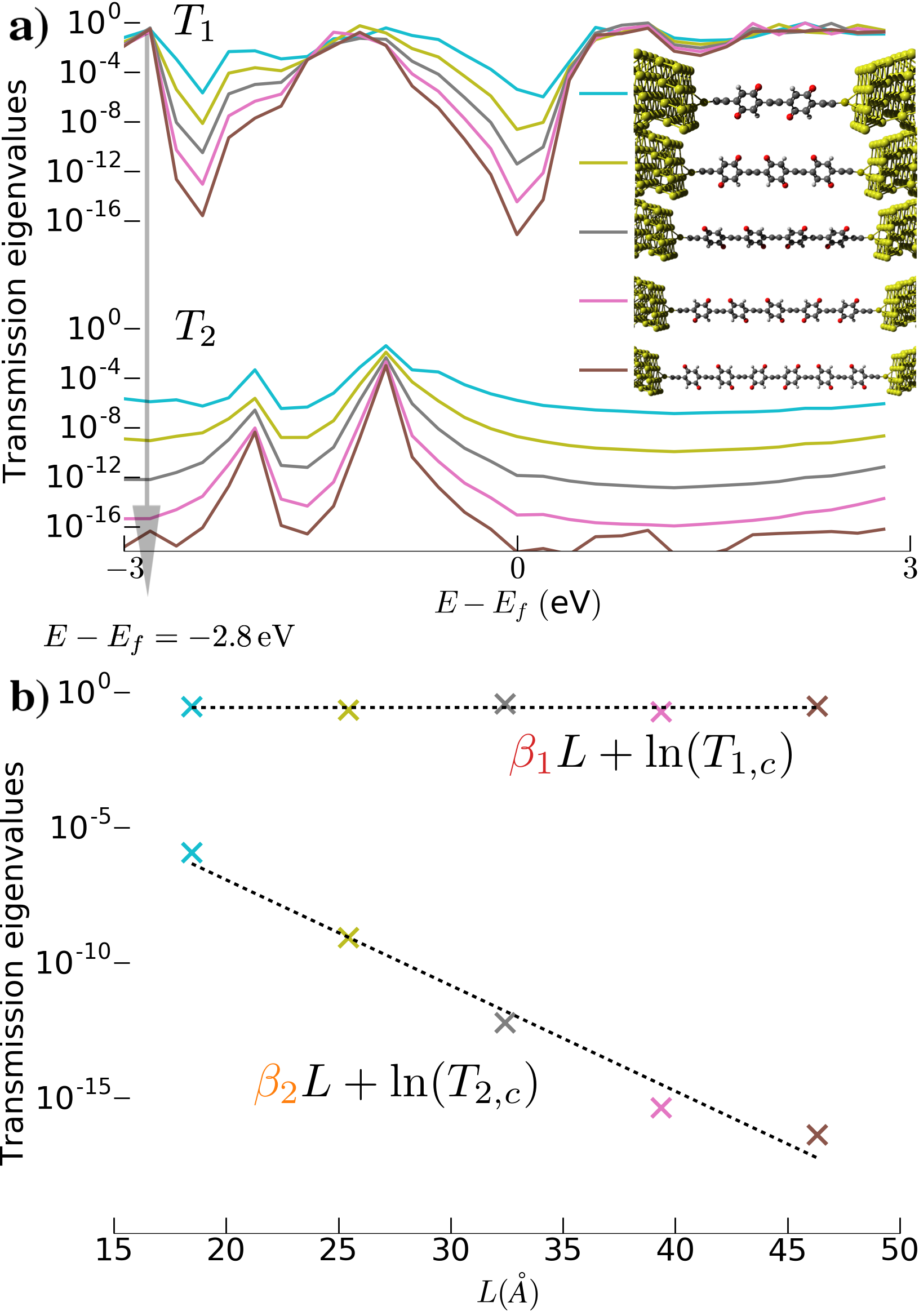}
	\caption{Example of a calculation comparing the decay constants $\beta$ estimated from a single complex band structure and six conventional transmission calculations. \textbf{a)} Quinone chains in junctions and the transmission of the two dominant transmission eigenvalue $T_1$ and $T_2$. \textbf{b)} Fits to estimate $\beta_n$ from the transmission functions in \textbf{a)}. In this example we show the fit at $E-E_f = -2.8$ eV.}
	\label{method}
\end{figure}

Here we initially show the calculated transmission eigenvalues for the quinone junction and outline the steps to calculate its length dependence. This leads us to the main result: the connection between the transmission length dependence of the two most dominant transmission eigenchannels ($\beta_1$ and $\beta_2$), and the decay of the Bloch states of the bridge material. Along the way, we discuss the effects of band gap size, destructive interference and eigenchannels with similar decay. 

The energy dependent transmission length dependence, $\beta_n(E)$, was found by fitting the transmission eigenvalues to the model,
\begin{align}
T_n(E) = T_{n,c} \, e^{-\beta_n(E) L},
\end{align}
where $L$ and $T_{n,c}$ are the bridge length and contact transmission eigenvalue, respectively. This is illustrated for $E-E_f=-2.8$ eV in Figure \ref{method} b). It is clear with this approach that the transmission eigenvalues do not always show a perfect exponential decay. This can be attributed to two effects. One, there a slight shift in the resonances for the different lengths, see $T_1$ at around $-1$ eV in Figure \ref{method} a). Two, at very low transmission we see numerically inaccuracy, see the brown curve of $T_2$ in Figure \ref{method} a). Despite this variation, a slope can be obtained for each channel as a function of energy, and from this a $\beta_n(E)$. 

\begin{figure}[h!]
 	\centering
	\includegraphics[width=0.99\linewidth]{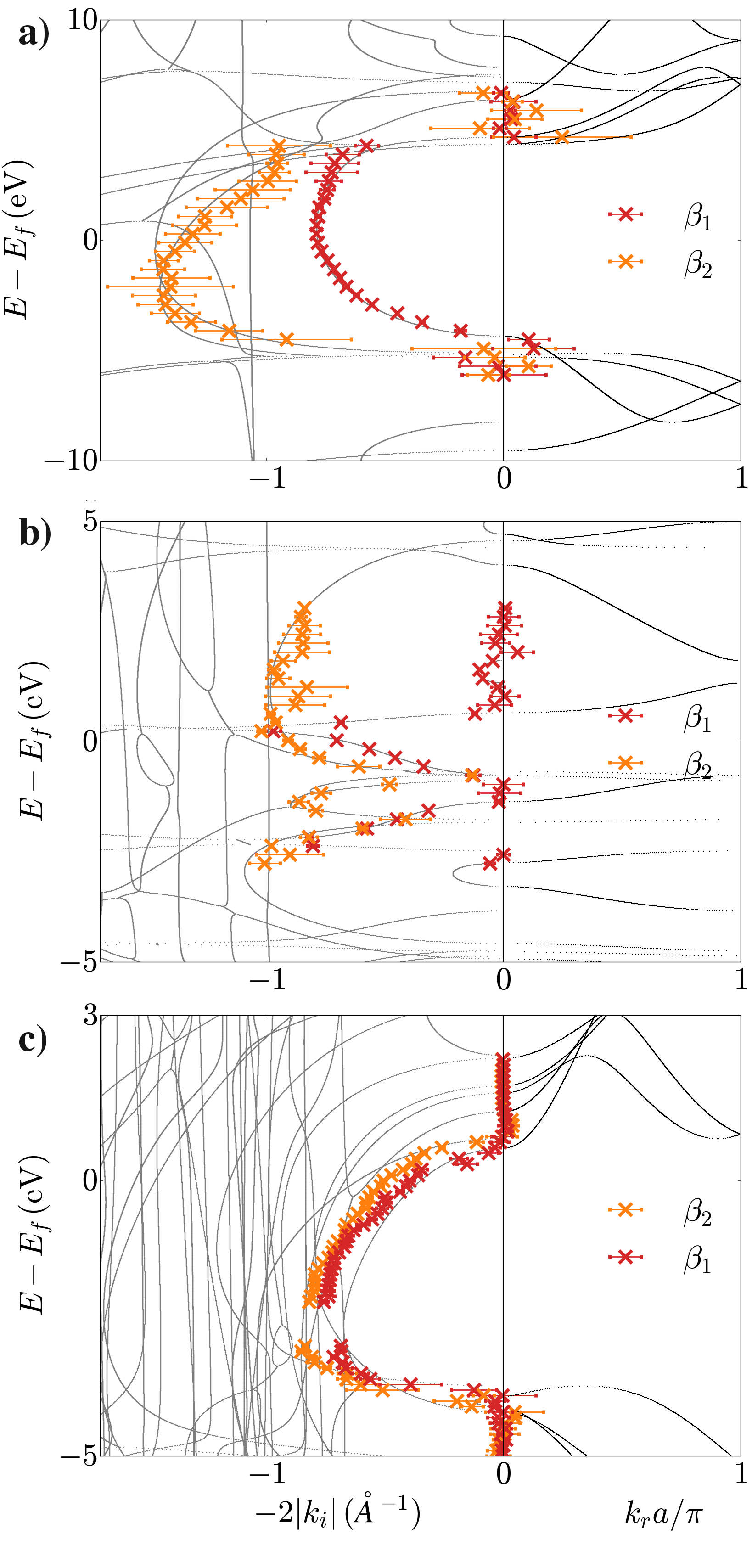}
	\caption{Complex band structure and corresponding $\beta$ values from transport calculations for three systems. \textbf{a)} Alkane chain. \textbf{b)} Quinone chain. \textbf{c)} Si - Si$_3$N$_4$ - Si }
	\label{res}
\end{figure}

Now we can directly show the connection between the length dependence of the transmission eigenvalues, $\beta_1$, $\beta_2$, and decay of the Bloch states. In Figure \ref{res} the estimated $\beta_n$-values and their standard deviations (errors from the fits) are plotted on top of the complex band structure of the bridge region. Note that the Fermi energy, $E_f$, is aligned in the two types of calculations. 

We briefly go through the general elements of the plots. On the right we have the black curves, which represent the real bands with a band gap around $E_f$. On the left we have the complex bands with $\beta_1$ and $\beta_2$ plotted on top. We note that the complex bands are plotted on $ 2 |k_i|$ scale, see Equation \ref{modeleq}. As with the model in Section \ref{tbsec}, we see bands that are connected to one or more real bands but not necessarily with the line shape of a half circle. Most of the small $|k_i|$ complex bands are connected to frontier real bands. The three systems show distinctly different trends, so we will discuss each one in turn.

The alkane chain in Figure \ref{res} a) has a large band gap, $\sim 8$ eV, characteristic of a saturated molecule. At the energies where real bands exist, the transmission goes to 1 and $\beta = 0$. The complex band with the smallest $|k_i|$ at the Fermi energy forms a distorted half-circle between the valence band and an unoccupied band. The length dependence of the first transmission eigenvalues, $\beta_1$, sits right on top of this band. Thus we conclude this band predicts the length dependence of most dominant transmission eigenvalue to a high degree. In other words the Bloch states and the eigenchannel (across the molecule) must have a very similar character.  $\beta_2$ has larger error but still sits on top of another, larger $|k_i|$ band. Interestingly, at the Fermi energy we see a complex band that is close to vertical and even though the Bloch states of this band has a slower decay they are not as relevant for transport. This shows an important fact, small $\beta$ or $|k_i|$ in a complex band is often, but not always, equivalent with a large contribution to transport.   

The quinone chain in Figure \ref{res} b) is a fully conjugated molecule with a small band gap $\sim 1$ eV; notice the difference in the energy scale. The symmetry of this molecule means that we have $\sigma$ and $\pi$ systems in the molecule, and consequently separable transmission and bands. Furthermore, this is a system that is known to be dominated by destructive quantum interference effects in the $\pi$ system near the Fermi level, see the sharp dip in $T_1$ near $E_f$ in Figure \ref{method} a). We see this manifest in the complex band structure as the line shape of the complex band $|k_i|$ is not a half circle but has a sharp feature around the Fermi energy, indicating a strong length dependence around a destructive interference. $\beta_1$ from the first eigenchannel follows this complex band within the gap. Also inside the band gap, we see $\beta_2$ following the second smallest $|k_i|$ complex band, similar to what we would expect for the $\sigma$ transmission. In summary, the quinone junction has shown us that a manifestation of destructive interference in transport can be seen inside the complex band structure as a complex band with a sharp feature.  

Finally, Figure \ref{res} c) shows silicon nitride, a system with a band gap of $\sim 4$ eV. Again we see that the complex band with the smallest $|k_i|$ at the Fermi energy forms a distorted half-circle. However, here the $\beta_1$ and $\beta_2$ do not sit right on top a particular band. Instead the transmission decays at a similar rate for the two channels. This could indicate the eigenchannels consist of a mix of Bloch states from different bands. However, other explanations are also possible. The small region without $\beta_1$ or $\beta_2$ values is where the band gap of the silicon electrodes suppresses transport through the junction. 

The three cases in Figure \ref{res} give a strong indication that the complex bands of the bridge material can be used to estimate the decay constant $\beta$ of the individual transmission eigenvalues. For the three types of junctions studied here, the first complex band and the calculated $\beta_1$ values match well with one another. $\beta_2$, however, only followed the complex band with the second smallest $|k_i|$ in two of the three cases. This leads us to another interesting point because one class of complex bands deserves particular attention: vertical bands. In the case of the alkane, this was the band that $\beta_2$ did not touch, but in the event that a band like this did contribute to transport, it would indicate transmission that is nearly energy independent. These bands could be reminiscent of ghost transmission \cite{Herrmann2010}, and suggest that these kinds of bands might manifest with small $\beta$ values when large basis sets are used.  

A final point to note here is that we have employed long molecular systems in the transport calculations to avoid effects related to the interfaces of the junctions (e.g. anchoring groups)\cite{peng2009conductance}. However, when we looked for these effects by including the short 1,2-ethanedithiol to estimate $\beta_n$ for the alkane junction, we only found small differences (see SI). Interestingly, this discussion of interface effects highlights that the physicists measure of choice, the complex band structure, provides the cleanest measure of the chemical influence of the backbone. At least at short backbone lengths, the fact that the binding groups are held constant might not be sufficient to remove their influence on the result obtained. 

\section{Conclusion}
To sum up we have presented a simple model junction that illustrated the connection between the Bloch states of each region in the junction with the junction scattering states. Extending to atomistic models, we found that the complex band structure of the bridge material can predict the length dependence of individual transmission eigenvalues to some degree, but the direct comparison has also revealed further details.
\begin{itemize}
\item The complex bands can exhibit many different forms, from the semi-circular shape seen in the model system calculations, to sharp dips in the case of destructive interference, to vertical, flat bands. 
\item There will be cases, for example the alkane, where transport is dominated by a single eigenchannel and the eigenchannel $\beta$ will track the complex band with the smallest $|k_i|$. But this is not necessarily general.
\item In the case where eigenchannel $\beta$s diverge from the complex bands, it appears that there may be a mixing of multiple bands into a single channel transmission, for example in Si$_3$N$_4$.
\item $|k_i|$ is not connected to the magnitude of the transmission. Despite the fact that the shape of the complex bands is reminiscent of the shape of the transmission, it is incorrect to infer that one predicts the other. The sharp dip in the smallest $|k_i|$ complex band for the quinone in particular has a strong resemblance to the $\pi$ transmission at this energy, but these dips represent distinct effects with the same underlying cause. 
\item Not every complex band contributes to the transmission. This was particularly clear in the case of the vertical band between the bands picked up by the first and second transmission channels for the alkane. In the cases explored here, the eigenchannel $\beta$s generally followed the smallest $|k_i|$ complex bands, but this need not be the case. \end{itemize}
The extent to which the properties of a central molecule or material will dominate in a heterogeneous conducting junction will certainly vary from junction to junction. The electrodes can have a significant influence, in terms of their density of states, their hybridization with the molecule/material, and in terms of their ability to perturb molecular/material energy levels through image-charge effects. Despite these factors, in the cases where the molecular/materials properties dominate, or can be engineered to do so, the complex band structure offers a useful window into the transport properties of the molecule/material alone. 

\begin{acknowledgments}
AJ, MS and GCS acknowledge financial support from the Danish Council for Independent Research, Natural Sciences and the Carlsberg Foundation. MGR was supported by startup funds from the Institute for Advanced Computational Science at Stony Brook University. 
\end{acknowledgments}

\bibliographystyle{apsrev4-1}
\bibliography{refs.bib}

\end{document}


\preprint{APS/123-QED}

\title{Supplementary Information}

\author{Anders Jensen}
\affiliation{Nano-Science Center and Department of Chemistry, University of Copenhagen, 2100 Copenhagen \O, Denmark}
\author{Mikkel Strange}
\affiliation{Nano-Science Center and Department of Chemistry, University of Copenhagen, 2100 Copenhagen \O, Denmark}
\affiliation{Department of Physics, Technical University of Denmark, 2800 Kongens Lyngby, Denmark}
\author{S\o ren Smidstrup}
\author{Kurt Stokbro}
\affiliation{QuantumWise A/S, Fruebjergvej 3, Box 4, 2100 Copenhagen, Denmark}
\author{Gemma C. Solomon}
\email{gsolomon@nano.ku.dk}
\affiliation{Nano-Science Center and Department of Chemistry, University of Copenhagen, 2100
Copenhagen \O, Denmark}
\author{Matthew G. Reuter}
\email{matthew.reuter@stonybrook.edu}
\affiliation{Department of Applied Mathematics \& Statistics and Institute for Advanced Computational Science, Stony Brook University, Stony Brook, New York 11794, United States}

\date{\today}
\maketitle

\section{Junction model}
\begin{figure}[h]
 	\centering
 	\includegraphics[width=0.5\linewidth]{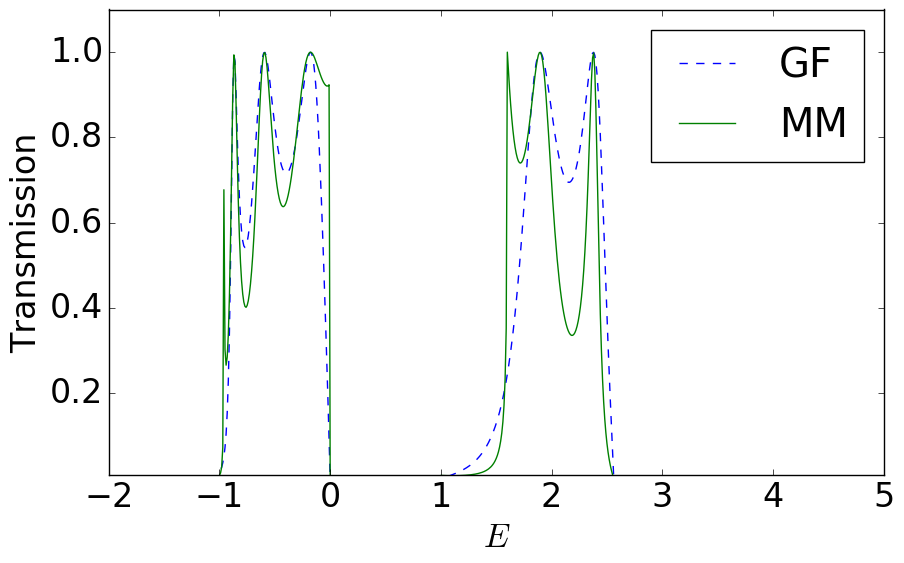}
 	\includegraphics[width=0.5\linewidth]{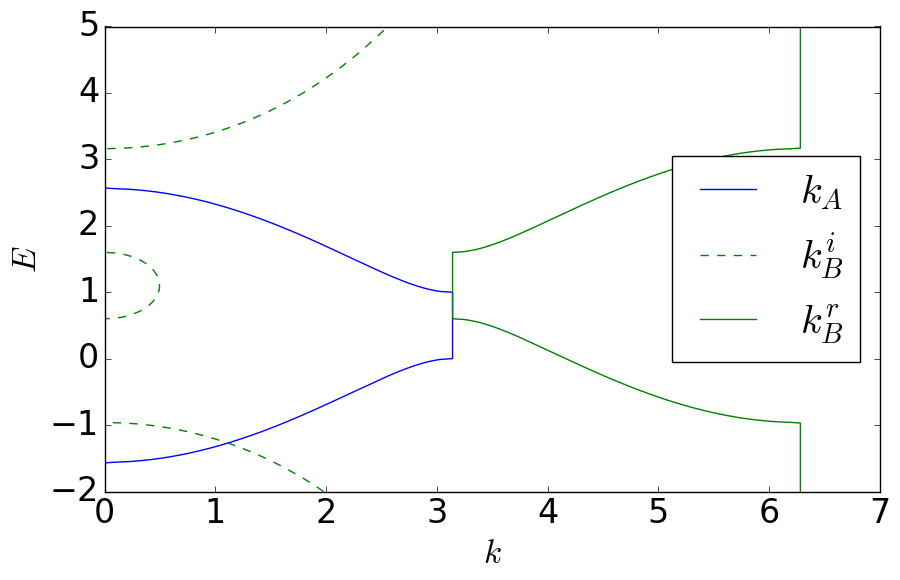}
 	\caption{Top) Transmission comparison of the mode matching method used in this paper vs. the standard Green's function approach. Bottom) Plot of energy as of function $k$-values}
	\label{tbmodel}
\end{figure}

Figure \ref{tbmodel} compares the transmission found using the expression derived in the paper using our the approximate method, mode matching (MM), and the standard Non-equlibrium Green's function (NEGF) approach.
The two methods share most peaks but are not completely identical. We find that the mode matching technique is highly dependent on the wave vectors ($k_A$ and $k_B$). Given the right combination of wave vectors we might find identical transmission.  

We also calculated the eigenvectors using NEGF approach as implemented in Atomic Simulation Environment (ASE). \cite{bahn}.
\begin{figure}[h]
 	\centering
 	\includegraphics[width=0.6\linewidth]{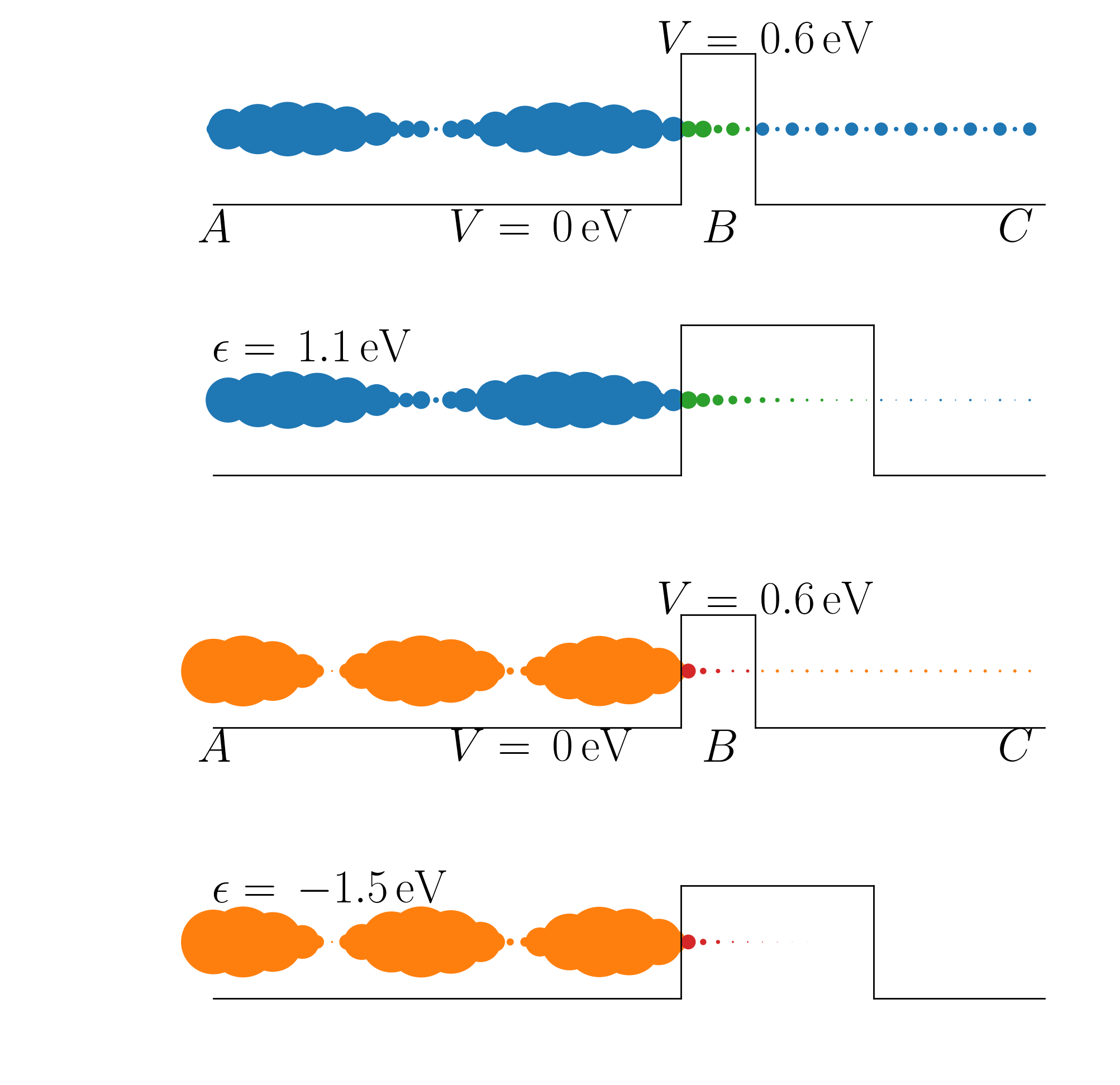}
 	\caption{Eigenchannels found using the NEGF method}
	\label{eig}
\end{figure}
We found that there are qualitative difference in the region A as we see large oscillations not present in the MM method. Note the region A is longer compared with Figure 3 in the paper to illustrate this point. The decay across to barrier in region B resembles the MM well and so does Region C. 

We conclude that our approximate MM approach gives a clear correlation between complex $k$ states and the corresponding transmission, however, we must remember that these states and their transmission is an approximation to the real junction states and their transmission.

\section{Short molecules and interface effects}
To investigate if interface effects gave poor estimates of $\beta_n$ we included the short 1,2-ethanedithiol which meant we now had six points to make the fit.
\begin{figure}[h]
 	\centering
 	\includegraphics[width=0.5\linewidth]{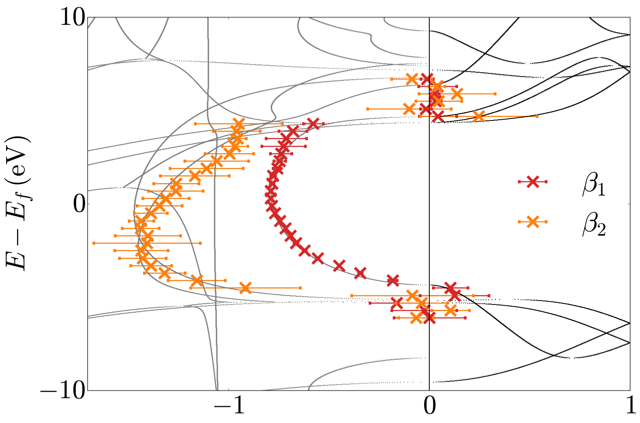}
 	\includegraphics[width=0.5\linewidth]{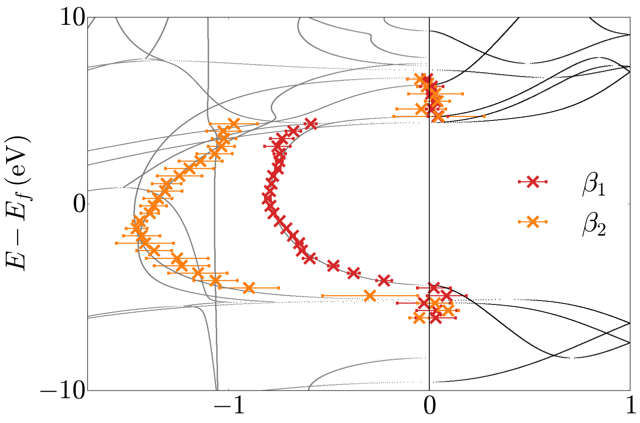}
 	\caption{Top) reprint of Figure 6 b) from the paper. Bottom) Now $\beta_n$ is estimated from 6 lengths including the short 1,2-ethanedithiol}
	\label{cbs}
\end{figure}
Generally we found smaller errors to our fit when we include the short molecule. However, we note a small kink in $\beta_1$ around $-2$ eV which could be interface effects.

\newpage
\bibliographystyle{apsrev4-1}
\bibliography{SIrefs.bib}